# Three-dimensional Integrated Guidance and Control for Leader-Follower Flexible Formation of Fixed Wing UAVs

**Praveen Kumar Ranjan**, Member, IEEE
Department of Electrical and Computer Engineering, The University of Texas at San Antonio, TX

**Abhinav Sinha**, Senior Member, IEEE
Department of Aerospace Engineering and Engineering Mechanics, University of Cincinnati, OH

**Yongcan Cao**, Senior Member, IEEE
Department of Electrical and Computer Engineering, The University of Texas at San Antonio, TX

*Abstract—* This paper presents a nonlinear integrated guidance and control (IGC) approach for flexible leader-follower formation flight of fixed-wing unmanned aerial vehicles (UAVs) while accounting for high-fidelity aerodynamics and thrust dynamics. Unlike conventional leader–follower schemes that fix the follower's position relative to the leader, the follower is steered to maintain range and bearing angles (which is the angle between its velocity vector and its line-of-sight (LOS) with respect to the leader) arbitrarily close to the prescribed values, enabling the follower to maintain formation on a hemispherical region behind the leader. The proposed IGC framework directly maps leader–follower relative range dynamics to throttle commands, and the follower's velocity orientation relative to the LOS to aerodynamic control surface deflections. This enables synergism between guidance and control subsystems. The control design uses a dynamic surface control–based backstepping approach to achieve convergence to the desired formation set, where Lyapunov barrier functions are incorporated to ensure the follower's bearing angle is constrained within specified bounds. Rigorous stability analysis guarantees uniform ultimate boundedness of all error states and strict constraint satisfaction in the presence of aerodynamic nonlinearities. The proposed flexible formation scheme allows the follower to have an orientation mismatch relative to the leader to execute anticipatory reconfiguration by transitioning between the relative positions in the admissible formation set when the leader aggressively maneuvers. The proposed IGC law relies only on relative information and onboard sensors without the information about the leader's maneuver, making it suitable for GPS-denied or non-cooperative scenarios. Finally, we present simulation results to vindicate the effectiveness and robustness of our approach.

## I. Introduction

Modern UAV missions increasingly demand coordinated group behavior, with formation flight providing the organizational structure needed for effective operation in tasks such as reconnaissance, interception, surveillance, target defense, and aerial refueling, e.g., see [1]–[5]. Formation control strategies are typically categorized as position-based, distance-based, or bearing-based, depending on the variables used to describe the desired geometry. Achieving and maintaining such formations requires coordination of flight variables, which can be obtained through onboard sensing [6], direct measurements [7], inter-vehicle communication [8], or state estimation [9]. Recent works highlight a shift toward formation control methods that achieve coordination without direct inter-vehicle communication, which helps alleviate network resources. In [10], the proposed control laws exploited relative measurements, where the follower was equipped with onboard sensors to gather observations and estimate the variables necessary to maintain formation. The authors in [11] proposed distance-based formation strategies based solely on relative position measurements in the local frame that were invariant to the orientation of the reference frame of each vehicle. In [12], formation was achieved through deviated pursuit guidance, requiring vehicles only to regulate the range and heading angle relative to the line-of-sight of the vehicles.

It is worth noting that most of the above-mentioned works constrain the agents in a rigid configuration relative to each other. In contrast, enabling flexibility in formation geometry allows agents to reconfigure is advantageous for practical deployment. To capture this idea, some studies have proposed time-varying formations, allowing agents to reshape into new configurations while maintaining robustness to challenges such as varying topology [13], intermittent communication [14], actuator failures [15], [16], and obstacle fields [17]. The authors in [18] allowed the formation to adapt, providing some degree of flexibility by permitting orientation mismatch between the agents. Although the above approaches allow some adaptability, the agents still remain constrained to fixed or rigid formations, highlighting the importance of formation strategies that can reshape dynamically in response to changing conditions. Therefore, the works in [19]–[22] introduced a novel notion of flexibility by allowing the follower to converge to a broader set of relative positions with respect to the leader. Specifically, in [19] the follower maintained formation by converging to a 2D semicircle behind the leader, whereas [20]–[22] extended this concept to a three-dimensional ring behind the leader.

Most of the existing formation strategies rely on guidance laws derived from simplified kinematic models, leaving the autopilots to handle the real-world dynamics. Such cascaded architecture often ignores actuator limitations and inner-loop delays, which can compromise performance during aggressive maneuvers or in high-disturbance environments. To address these shortcomings, the concept of integrated guidance and control (IGC) aims at merging mission-level command generation with vehicle actuation into a unified design. Although explored in interceptor systems [23]–[25], the application of IGC to fixed-wing UAVs has received limited attention. In [26], a partial IGC law was developed for the formation flight of a fixed-wing UAV to synthesize both guidance and control in one loop, thereby minimizing the overall time lag to the best possible extent. The authors in [27] addressed waypoint navigation of fixed-wing UAVs, proposing an IGC-based approach by linearizing the nonlinear model around the equilibrium points. The authors in [28] proposed IGC laws for fixed-wing path following by decoupling the tightly coupled longitudinal–lateral dynamics to achieve tractable controller synthesis. In [29], a 3D-IGC law was proposed for bank-to-turn aircraft for target interception using a robust dynamic inversion–based backstepping approach enhanced with dynamic surface control. An adaptive IGC framework was developed in [30] for automatic landing of fixed-wing UAVs, where online neural networks compensated for modeling uncertainties.

This work is motivated by the need to endow fixed-wing UAV formations with greater maneuvering flexibility and robustness in realistic 3D flight regimes by allowing the follower to maintain formation within a set of admissible positions rather than a single point. Such flexibility allows the follower to anticipate the leader's motion by adjusting its trajectory (taking sharper, larger turns or climbing/descending with respect to the leader) to reduce maneuvering efforts while maintaining situational awareness and tactical advantage. These adaptive maneuvers are particularly critical in air-to-air combat and other tactical, reconnaissance, or strategic missions, where the ability to reconfigure relative positioning can provide decisive operational benefits. Conventional formation strategies typically utilize cascaded guidance and control architectures that neglect the feasibility of maneuver execution by the follower. In practice, fixed-wing UAVs are subject to aerodynamic coupling, actuator limits, and inner-loop dynamics that significantly affect their maneuvering capabilities. Ignoring these factors can result in degraded performance, especially during aggressive maneuvers or in contested environments where precision and responsiveness are critical. Motivated by these challenges, we develop 3D relational maneuvering strategies that provide followers with the flexibility to adapt their relative positioning while explicitly accounting for the UAV dynamics through an IGC design. The main contributions of the paper are as follows:

- We propose a 3D flexible leader-follower formation scheme that allows the follower to converge to a hemisphere behind the leader, maintaining tactical advantage over the leader. This extends our previously introduced novel notion of flexibility (follower restricted to a 2D semicircle [19] and a 3D ring [20]–[22]) to a broader set, enabling the follower to better anticipate the leader and execute appropriate maneuvers.
- We propose an IGC law for the follower to maintain the proposed flexible formation, explicitly incorporating the aerodynamic variations into the vehicle dynamics. Unlike conventional IGC methods for fixed-wing UAVs [26]–[29], our approach incorporates aerodynamic surface dynamics and the nonlinear propeller–motor dynamics, bridging the gap between theoretical control design and practical UAV implementation.
- The proposed approach incorporates bearing angle constraints via Lyapunov barrier function in order to ensure that the follower remains confined within a 3D conical region behind the leader. Unlike prior formation control methods [6]–[11] that neglect state constraints, the integration of barrier functions into the backstepping/dynamic surface control framework simultaneously guarantees constraint satisfaction and closed-loop stability.
- Unlike previous works [8], [9], [14], [16], [21], the proposed scheme is realized through relative information for formation along with standard onboard sensing such as airspeed and inertial measurement required for the IGC design. This reduces the complexity of the leader-follower state-space, lowers computational effort, and avoids reliance on explicit vehicle-to-vehicle communication, making our approach effective in GPS-denied and communication-degraded environments.

## II. Problem Formulation

In this section, we formulate the 3D leader-follower relative kinematics and provide a complete 6-DOF model of a fixed-wing aircraft. Based on these models, we then establish the control objectives for the proposed flexible formation strategy.

### A. Leader-Follower Relative Kinematics

Consider the 3D relative geometry between two UAVs, namely, a leader and a follower, in the inertial frame, denoted by mutually orthogonal $X_I, Y_I, Z_I$ axes. The $i^{\text{th}}$ UAV moves at a speed $V_i$ such that its velocity subtends a flight path angle $\gamma_i \in (-\pi/2, \pi/2)$ to the inertial $X_I - Y_I$ plane and heading angle $\chi_i \in (-\pi, \pi)$ with respect to the inertial $X_I$ axis. The subscript $i \in \{l, f\}$ represents the leader and follower-related variables, respectively. Assuming the UAVs to be non-holonomic agents, their

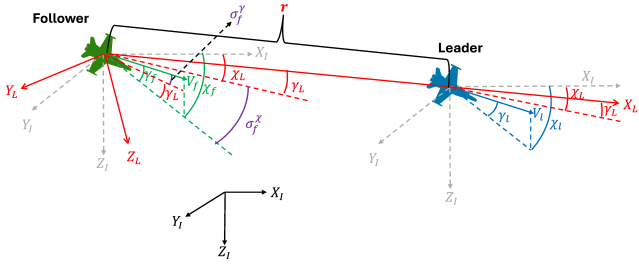

Fig. 1: Leader-follower relative geometry.

kinematics are governed by

$$\dot{x}_i = V_i \cos \gamma_i \cos \chi_i, \quad \dot{y}_i = V_i \cos \gamma_i \sin \chi_i, \quad \dot{z}_i = -V_i \sin \gamma_i, \tag{1}$$

where $\mathbf{p}_i = [x_i, y_i, z_i]^\top$ denotes the instantaneous position of the $i^{\text{th}}$ UAV in the inertial frame of reference. In Fig. 1, the leader and the follower UAVs are separated by a distance $r$, whereas the line-of-sight (LOS) from the follower to the leader subtends $\gamma_L \in (-\pi/2, \pi/2)$ (denoting elevation LOS angle) from the $X_I - Y_I$ plane and sweeps $\chi_L \in [-\pi, \pi)$ (azimuth LOS angle) from the $X_I$ axis. Therefore, we have an LOS frame denoted by mutually orthogonal $X_L, Y_L, Z_L$ axes, which has $X_L$ aligned along the LOS. The equations of relative motion between the leader and the follower in such a scenario are given as

$$\dot{r} = V_l \left( \sin \gamma_L \sin \gamma_l + \cos \gamma_L \cos \gamma_l \cos (\chi_L - \chi_l) \right) \\ - V_f \left( \sin \gamma_L \sin \gamma_f + \cos \gamma_L \cos \gamma_f \cos (\chi_L - \chi_f) \right), \tag{2}$$

$$r\dot{\gamma}_L = V_l \left( \cos \gamma_L \sin \gamma_l - \cos \gamma_l \sin \gamma_L \cos (\chi_L - \chi_l) \right) \\ - V_f \left( \cos \gamma_L \sin \gamma_f - \cos \gamma_f \sin \gamma_L \cos (\chi_L - \chi_f) \right), \tag{3}$$

$$\dot{\chi}_L = \frac{-V_l \cos \gamma_l \sin (\chi_L - \chi_l) + V_f \cos \gamma_f \sin (\chi_L - \chi_f)}{r \cos \gamma_L}. \tag{4}$$

*Remark* 1 The above equations are obtained by projecting the relative velocity between the leader and the follower in the inertial frame of reference (as given in (1)) on the LOS frame of reference. This is unlike the usual convention, as seen in, e.g., [31].

*Remark* 2 We choose to transform the relative kinematics between the leader and the follower in spherical coordinates about the leader. Such considerations allow our control design to rely only on relative information, circumventing the need for vehicle-to-vehicle or vehicle-to-ground station communication.

*Definition* 1 We define the elevation and the azimuth bearing angles, $\sigma_i^\gamma$, and $\sigma_i^\chi$, respectively, as the angles that the vehicle velocity vector subtends with the LOS in the inertial frame of reference, as seen in Fig. 1. Mathematically, these angles are given as,

$$\sigma_i^\gamma = \gamma_i - \gamma_L, \quad \sigma_i^\chi = \chi_i - \chi_L. \tag{5}$$

*Definition* 2 We define the 3D bearing angle (dependent on $\sigma_i^\gamma$ and $\sigma_i^\chi$) as the angle between the UAV's velocity vector and the LOS, given as,

$$\cos \sigma_i = \sin \gamma_L \sin \gamma_i + \cos \gamma_L \cos \gamma_i \cos (\chi_L - \chi_i). \tag{6}$$

Note that when both $\sigma_i^\gamma \to 0$ and $\sigma_i^\chi \to 0$ or $\sigma_i \to 0$, the UAV's velocity is perfectly aligned along the LOS.

*Definition* 3 (Notion of staying ahead/behind the leader) We say that the follower is behind the leader if $\sigma_l \in [0, \pi/2)$, otherwise it is ahead of the leader when $\sigma_l \in [\pi/2, \pi]$.

*Remark* 3 The concept of staying behind the leader is fundamental in designing maneuvering strategies for flexible geometry formations. Specifically, staying behind the leader provides the follower with tactical advantages such as improved situational awareness of the leader's maneuvers and greater flexibility to adjust its own trajectory without risking collision or loss of formation. Assumption 3 further formalizes this requirement by imposing mild conditions on the leader's trajectory, thereby guaranteeing that the leader's velocity never points toward the follower.

### B. Nonlinear Dynamics of a Fixed-Wing UAV

We now present a comprehensive mathematical model with aerodynamic and thrust characterization to capture the flight behavior of a fixed-wing UAV. Assuming the aircraft is a rigid body and Earth is flat, combined with translational kinematic equations (1), the complete 6-DOF equations of motion for the follower UAV are given as [26], [32], [33]

$$\dot{V}_f = \frac{F_y s_{\beta_f} - F_D c_{\beta_f} + T_f c_{\alpha_f} c_{\beta_f}}{m_g} - g s_{\gamma_f}, \tag{7}$$

$$\dot{\gamma}_f = -\frac{g c_{\gamma_f}}{V_f} + \frac{T_f (c_{\mu_f} s_{\alpha_f} + c_{\alpha_f} s_{\beta_f} s_{\mu_f})}{m_g V_f} \\ - \frac{F_Y c_{\beta_f} s_{\mu_f}}{m_g V_f} + \frac{F_L c_{\mu_f}}{m_g V_f} - \frac{F_D s_{\beta_f} s_{\mu_f}}{m_g V_f}, \tag{8}$$

$$\dot{\chi}_f = \frac{T_f (s_{\alpha_f} s_{\mu_f} - c_{\alpha_f} c_{\mu_f} s_{\beta_f})}{m_f V_f c_{\gamma_f}} + \frac{F_Y c_{\beta_f} c_{\mu_f}}{m_g V_f c_{\gamma_f}} \\ + \frac{F_L s_{\mu_f}}{m_g V_f c_{\gamma_f}} + \frac{F_D c_{\mu_f} s_{\beta_f}}{m_f V_f c_{\gamma_f}}, \tag{9}$$

$$\dot{\mu}_f = \frac{p c_{\alpha_f}}{c_{\beta_f}} + \frac{r s_{\alpha_f}}{c_{\beta_f}} - \frac{g c_{\mu_f} c_{\gamma_f} \tan \beta_f}{V_f} + \frac{F_Y C_3}{m_g V_f} \\ + \frac{-F_D c_{\alpha_f} + F_L s_{\alpha_f}}{m V_f} C_1 + \frac{F_L c_{\alpha_f} + F_D s_{\alpha_f}}{m_g V_f} C_2, \tag{10}$$

$$\dot{\alpha}_f = -\frac{F_L + T_f s_{\alpha_f} - m_g g c_{\gamma_f} c_{\mu_f}}{m_g V_f c_{\beta_f}} \\ - p c_{\alpha_f} \tan \beta_f + q - r s_{\alpha_f} \tan \beta_f, \tag{11}$$

$$\dot{\beta}_f = \frac{F_y c_{\beta_f} + F_D s_{\beta_f} - T_f c_{\alpha_f} s_{\beta_f}}{m_i V_i} \\ + \frac{g c_{\gamma_f} s_{\mu_f}}{V_f} - r c_{\alpha_f} + p s_{\alpha_f}, \tag{12}$$

:                                                                                                                             3

$$\begin{bmatrix} \dot{p} \\ \dot{q} \\ \dot{r} \end{bmatrix} = \begin{bmatrix} \Gamma_1 pq - \Gamma_2 qr + \Gamma_3 \left( l - Q_P \right) + \Gamma_4 n \\ \Gamma_5 pr - \Gamma_6 \left( p^2 - r^2 \right) + \frac{1}{J_y} m \\ \Gamma_7 pq - \Gamma_1 qr + \Gamma_4 \left( l - Q_P \right) + \Gamma_8 n \end{bmatrix}, \quad (13)$$

where we use the shorthand notations $c_{(\cdot)} = \cos(\cdot)$ and $s_{(\cdot)} = \sin(\cdot)$. Note that, for brevity, we have dropped the subscript for the follower here since we will be dealing with the dynamics of only one aircraft. The UAV has mass $m_g$ and is subject to gravitational acceleration $g$. The aerodynamic angles $\alpha_f, \beta_f, \mu_f$ denote the angle of attack, side slip angle, and velocity vector roll angle, respectively. The angular rates are denoted by $p$, $q$, and $r$, corresponding to roll, pitch, and yaw rates, respectively. The aerodynamic forces are $F_L$ (lift), $F_D$ (drag), $F_Y$ (side force), and $T_f$ (rotor thrust). Similarly, the aerodynamic moments are $l$ (roll torque), $m$ (pitch torque), and $n$ (yaw torque). The propeller torque is represented by $Q_P$, while $\Gamma_{(\cdot)}$ denotes constants related to the UAV's moments of inertia. For compactness, we also define the auxiliary terms

$$C_1 = s_{\alpha_f} \tan \beta_f + \tan \gamma_f (s_{\alpha_f} s_{\mu_f} - c_{\alpha_f} s_{\beta_f} c_{\mu_f}), \quad (14)$$
$$C_2 = s_{\alpha_f} \tan \beta_f + \tan \gamma_f (s_{\alpha_f} s_{\mu_f} - c_{\alpha_f} s_{\beta_f} c_{\mu_f}), \quad (15)$$
$$C_3 = c_{\beta_f} c_{\mu_f} \tan \gamma_f. \quad (16)$$

The aerodynamic forces and torques acting on the follower UAV are modeled as,

$$F_L = Q \left[ C_{L0} + C_{L\alpha} \alpha_f + C_{Lq} \frac{c}{2V_f} q + C_{L\delta_e} \delta_e \right] \quad (17)$$

$$F_D = Q \left[ C_{D0} + C_{D\alpha} \alpha_f + C_{Dq} \frac{c}{2V_f} q + C_{D\delta_e} \delta_e \right] \quad (18)$$

$$F_Y = Q \Big[ C_{Y0} + C_{Y\beta} \beta_f + \frac{C_{Yp} c}{2V_f} p + \frac{C_{Yr} b}{2V_i} r$$
$$+ C_{Y\delta_a} \delta_a + C_{Y\delta_r} \delta_r \Big], \quad (19)$$

$$l = Qb \Big[ C_{l0} + C_{l\beta} \beta + C_{lp} \frac{c}{2V_f} p + C_{lr} \frac{b}{2V_f} r$$
$$+ C_{l\delta_a} \delta_a + C_{l\delta_r} \delta_r \Big], \quad (20)$$

$$m = Qc \left[ C_{m0} + C_{m\alpha} \alpha_f + C_{mq} \frac{c}{2V} q + C_{m\delta_e} \delta_e \right], \quad (21)$$

$$n = Qb \Big[ C_{n0} + C_{n\beta} \beta + C_{np} \frac{c}{2V} p$$
$$+ C_{nr} \frac{b}{2V} r + C_{n\delta_a} \delta_a + C_{n\delta_r} \delta_r \Big], \quad (22)$$

where we denote the air density by $\rho$, the effective wing area by $S$, and the dynamic pressure by $Q = \frac{1}{2}\rho V_f^2 S$. The wing chord length and span are represented by $c$ and $b$, respectively. The control surface deflections are given by $\delta_a$ (aileron), $\delta_e$ (elevator), and $\delta_r$ (rudder). Quantities $C_{(\cdot)}$ denote the aerodynamic coefficients corresponding to these parameters. The convention of control surface deflections along with aerodynamic forces and moments acting on the fixed-wing UAV are depicted in Fig. 2, where mutually orthogonal axes $x_b, y_b, z_b$ denote the body-fixed reference frame.

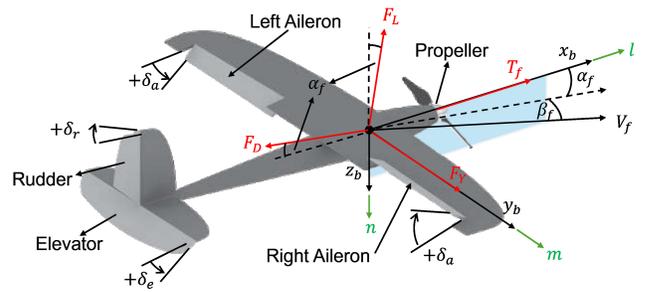

Fig. 2: Fixed-Wing UAV dynamics

Based on the propeller theory, the thrust and torque produced by the propeller are given by [33],

$$T_f = \left( \frac{\rho D^4 C_{T_0}}{4\pi^2} \right) \Omega_P^2 + \left( \frac{\rho D^3 C_{T_1} V_f}{2\pi} \right) \Omega_P + \rho D^2 C_{T_2} V_f^2, \quad (23)$$

$$Q_P = \left( \frac{\rho D^5 C_{Q_0}}{4\pi^2} \right) \Omega_P^2 + \left( \frac{\rho D^4 C_{Q_1} V_f}{2\pi} \right) \Omega_P + \rho D^3 C_{Q_2} V_f^2, \quad (24)$$

where we denote $C_{T_0}$, $C_{T_1}$, $C_{T_2}$ as the propeller thrust coefficients, and $C_{Q_0}$, $C_{Q_1}$, and $C_{Q_2}$ as the propeller torque coefficients. The rotor angular speed is represented by $\Omega_P$. Additionally, the rotor RPM is modeled as a first-order dynamics governed by the following equation,

$$\dot{\Omega}_P = \frac{K_Q \left( (V_{\max} \delta_t - K_V \Omega_P) - i_0 \right)}{R J_P} - \frac{Q_P}{J_P}, \quad (25)$$

where $K_V$, $K_Q$ are motor related constants, $R$ is the motor electric resistance, $V_{\max}$ is the maximum voltage, $i_0$ is the motor currents, $J_P$ is the rotor motor combined moment of inertia and $\delta_t$ is the throttle command.

### C. Design of the Control Objectives

In this paper, we are concerned with the relational maneuvering of UAVs for leader-follower formation, where instead of the follower converging to a fixed position relative to the leader, it can occupy multiple positions and still remain in formation with the leader without losing cohesion. To this end, we propose a flexible relational maneuvering strategy that allows the follower to maintain a fixed distance $r_d$, a fixed desired elevation bearing angle $\sigma_d^\gamma$, and a fixed desired azimuth bearing angle $\sigma_d^\chi$ with respect to the leader.

*Definition* 4 (Fixed Elevation-Azimuth Maneuver, or FEAM) The formation maneuver FEAM, $\mathscr{F}$, is the set of the follower's positions that satisfy the distance and bearing-angle constraints

$$\mathscr{F} = \left\{ \mathbf{p}_f \in \mathbb{R}^3 \mid \|r\| = r_d, \ \sigma_f^\gamma = \sigma_{fd}^\gamma, \ \sigma_f^\chi = \sigma_{fd}^\chi \right\}, \quad (26)$$

where $\mathbf{p}_f$ denotes the follower's position vectors in the inertial frame.

*Remark 4* Fixing all the desired values specifies a single point on the sphere of radius $r_d$ centered at the leader. If instead only a subset of the constraints is enforced, flexibility in maneuver could be assured. An important advantage of FEAM lies in prescribing only the desired range, yielding a continuous set of admissible follower positions on a sphere of radius $r_d$ around the leader. The desired bearing angles act as velocity orientation constraints rather than absolute position specifications, thereby affording the follower the flexibility to maneuver and reconfigure within this admissible set while preserving the intended formation geometry.

In particular, fixing $\sigma_f^\gamma$ and $\sigma_f^\chi$ irrespective of the LOS angle or the leader's bearing angle will lead to the follower converging on a sphere of fixed radius $r_d$ around the leader. In some scenarios, it may even go ahead of the leader, which may not be tactically advantageous because it could obstruct the follower's visibility. To avoid this, we impose explicit bounds on the follower's bearing angles so that it remains confined behind the leader, by maneuvering in such a way that guarantees $\sigma_l \in (-\pi/2, \pi/2)$ (according to Definition 3).

Fig. 3 demonstrates the leader-follower formation under FEAM, when both the agents' velocities are aligned along the inertial $X$-axis, with the hemisphere indicating the admissible locations for the follower to stay behind the leader. In Fig. 3, the solid purple lines represent the latitude and longitude on the hemisphere dependent upon the desired bearing angles, while the dotted purple arcs depict the constrained subset of the hemisphere to which the follower may converge, depending on the imposed bearing angle constraints. In this work, we adopt an IGC framework, wherein the guidance and control loops are synthesized in a unified manner rather than through a sequential design. Therefore, instead of designing the follower's acceleration components, we directly aim to design the low-level commands, which will ensure tighter coupling between guidance and control subsystems.

We are now in a position to state the main problem addressed in this work. The objective of this work is to design the follower's control inputs ($\mathbf{u}_f = [\delta_t, \delta_a, \delta_e, \delta_r]^\top$) to attain FEAM by ensuring, i) $\lim_{t\to\infty} r \to r_d$, ii) $\lim_{t\to\infty} \sigma_f^\gamma \to \sigma_{fd}^\gamma$, and iii) $\lim_{t\to\infty} \sigma_f^\chi \to \sigma_{fd}^\chi$, while ensuring $\sigma_l \in (-\pi/2, \pi/2), \forall t \geq 0$. Before proceeding with the control design, we provide an important result related to the Lyapunov barrier function, a control-theoretic technique utilized for ensuring constraints on the desired state or output variables.

LEMMA 1 ( [34]) *Consider the open sets, $\mathcal{Z}_1 \coloneqq \{z_1 \in \mathbb{R}, -a < z_1 < b\} \subset \mathbb{R}$, $\mathcal{N} \coloneqq \mathbb{R} \times \mathcal{Z}_1 \subset \mathbb{R}^2$ and the system $\dot{\eta} = h(t, \eta)$, where $\eta \coloneqq [w, z_1]^\top \in \mathcal{N}$ and $h$ is piecewise continuous in $t$ and locally Lipschitz in $z$, uniform in $t$, such that $h: \mathbb{R}_{>0} \times \mathcal{N} \to \mathbb{R}^2$. Suppose there exist two continuously differentiable and positive definite functions $U: \mathbb{R} \to \mathbb{R}_{\geq 0}$ and $V_1$, such that $V_1(z_1) \to \infty$, as $z_1 \to -a$ or $z_1 \to b$ and $\gamma_1(\|w\|) \leq U(\|w\|) \leq \gamma_2(\|w\|)$, where $\gamma_1$ and $\gamma_2$ are $\mathcal{K}_\infty$ functions. Let $V_2(\eta) = V_1(z_1) +$*

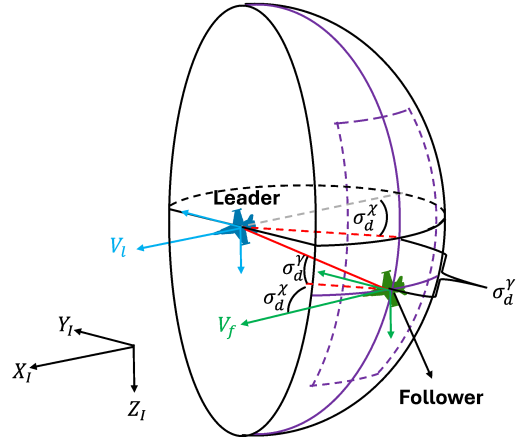

Fig. 3: Illustration of the FEAM scheme.

*$U(w)$ and $\epsilon(0)$ belong to the set $\epsilon \in (-a, b)$. If the given inequality holds, $\dot{V} = \frac{\partial V}{\partial \eta} h \leq 0$, then $\epsilon(t)$ remains in the open set $\epsilon \in (-a, b), \forall t \in [0, \infty)$.*

## III. Main results

In this section, we first design the throttle commands to regulate the follower's distance to the leader, and subsequently develop a bearing-angle controller that shapes the follower's velocity orientation via aerodynamic control surfaces while ensuring constraint satisfaction and stability.

### A. Design of the Range Controller

We define the range error as $e_r = r - r_d$. Differentiating $e_r$ with respect to time and using (2), we obtain the dynamics of the range error as

$$\dot{e}_r = V_l \left(\sin\gamma_L \sin\gamma_l + \cos\gamma_L \cos\gamma_l \cos(\chi_L - \chi_l)\right) \\ - V_f \left(\sin\gamma_L \sin\gamma_f + \cos\gamma_L \cos\gamma_f \cos(\chi_L - \chi_f)\right), \quad (27)$$

since $\dot{r}_d = 0$. Letting $x_1^r = V_f, x_2^{r*} = T_f, x_2^r = \Omega_P$ and $u_r = \delta_t$ and using (7), (23), (24), (25), and (27), we can express the relationship between the range error and the throttle command in a strict feedback-like form as

$$\dot{e}_r = f_0 + g_0 x_1^r + d_0^r \quad (28a)$$
$$\dot{x}_1 = f_1^r + g_1^r x_2^{r*} + d_1^r \quad (28b)$$
$$\dot{x}_2^r = f_2^r + g_2^r u_r + d_2^r \quad (28c)$$

where $f_0^r = V_l \left(\sin\gamma_L \sin\gamma_l + \cos\gamma_L \cos\gamma_l \cos(\chi_L - \chi_l)\right)$, $g_0^r = -\left(\sin\gamma_L \sin\gamma_f + \cos\gamma_L \cos\gamma_f \cos(\chi_L - \chi_f)\right)$, $f_1^r = \frac{F_y \sin\beta - F_D \cos\beta}{m} - g\sin\gamma$, $g_1^r = \frac{\cos\alpha \cos\beta}{m}$, $f_2^r = -\frac{K_Q K_V^m}{RJ_P} - \frac{K_Q i_0}{J_P} - \frac{1}{J_P}(b_0 \Omega_P^2 + b_1 \Omega_P + b_2)$, $g_2^r = \frac{K_Q}{RJ_P}$ and $d_i^r \ \forall \ i \in \{0, 1, 2\}$ denote the residual dynamics. We assume that the model uncertainties are unknown but bounded, that is, $|d_i^r| < \overline{d}_i^r \ \forall \ i \in \{0, 1, 2\}$.

Here, we utilize the method of Dynamic Surface Control [35], which employs first-order filters to avoid repeated differentiation of the virtual control laws to enable smooth and implementable control design while retaining rigorous stability guarantees. To this end, we define the errors between the desired and the filtered virtual control variables as

$$\tilde{x}_1^r = x_{1c}^r - x_{1d}^r, \quad \tilde{x}_2^r = x_{2c}^r - x_{2d}^r, \quad (29)$$

where $x_{1d}^r = V_{f_d}$ denotes the desired speed, $x_{2d}^r = \Omega_{f_d}$ denotes the desired propeller rotational speed, $x_{1c}^r = V_{f_c}$ denotes the filtered desired speed and $x_{2c}^r = \Omega_{f_c}$ denotes the filtered value of the desired propeller speed. Additionally, we define the error between the actual and filtered values of these variables as,

$$s_1^r = x_1^r - x_{1c}^r, \quad s_2^r = x_2^r - x_{2c}^r. \quad (30)$$

*Remark 5* Note that $x_2^{r*}$ is a nonlinear function of the inner state $x_2^r$. Therefore, we employ the mean value theorem to linearly approximate the deviation between the actual thrust and the desired thrust in terms of differences between the actual and the desired propeller speed.

Specifically, we obtain,

$$x_2^{r*} - x_{2d}^{r*} = (2A\tilde{\Omega} + B)(x_2^r - x_{2d}^r) = P_r(x_2^r - x_{2d}^r), \quad (31)$$

where $\tilde{\Omega} = x_{2d}^r + \eta_\Omega(x_2^r - x_{2d}^r)$ for some $\eta_\Omega \in (0,1)$. On substituting (31) in (28) and using eqs. (29) and (30), the error dynamics can now be expressed as,

$$\begin{aligned}\dot{e}_r &= f_0^r + g_0^r x_{1d}^r + g_0^r(s_1^r + \tilde{x}_1^r) + d_0^r, \\ \dot{x}_1^r &= f_1^r + g_1^r x_{2d}^r + g_1^r P_r(s_2^r + \tilde{x}_2^r) + d_1^r.\end{aligned} \quad (32)$$

Based on dynamic surface control, the proposed throttle control law can be given as

$$\delta_t = u_r = \frac{-f_2^r - K_2^r s_2^r - k_2^r \text{sign}(s_2^r) + \dot{x}_{2c}^r}{g_2^r}, \quad (33)$$

where one has the terms

$$x_{1d}^r = \frac{-f_1 - K_0^r e_r - k_0^r \text{sign}(e_r)}{g_0^r}, \quad (34a)$$

$$\tau_1^r \dot{x}_{1c}^r = x_{1d}^r - x_{1c}^r, \quad x_{1c}^r(0) = x_{1d}^r(0), \quad (34b)$$

$$x_{2d}^{r*} = \frac{-f_1^r - K_1^r s_1^r - k_1^r \text{sign}(s_1^r) + \dot{x}_{1c}^r}{g_1^r}, \quad (34c)$$

$$x_{2d}^r = \frac{-B + \sqrt{B^2 - 4A(C - x_{2d}^{r*})}}{2A}, \quad (34d)$$

$$\tau_2^r \dot{x}_{2c}^r = x_{2d}^r - x_{2c}^r, \quad x_{2c}^r(0) = x_{2d}^r(0), \quad (34e)$$

where $K_0^r, k_0^r, K_1^r, k_1^r, K_2^r, k_2^r$ are the controller gains, and $\tau_1^r, \tau_2^r$ are the filter constants to be designed subsequently. Further, (34d) provides the nonlinear mapping function between thrust and the propeller speed, obtained by inverting the thrust model in (23), where $A = \frac{\rho D^4 C_{T_0}}{4\pi^2}$, $B = \frac{\rho D^3 C_{T_1} V_f}{2\pi}$ and $C = \rho D^2 C_{T_2} V_f^2$.

*Assumption 1* We assume that the time derivatives of the follower's desired speed and the propeller speed are bounded, that is, $|\dot{x}_{id}^r| < \epsilon_i^r$, $\forall i \in \{1,2\}$, where $\epsilon_i^r > 0$ is a constant.

*Remark 6* Such an assumption is routinely adopted in dynamic surface control-based approaches to guarantee that filtering virtual controls are well-defined and smooth. From a practical standpoint, it is reasonable since the dynamics of both airspeed and propeller rotation are inherently constrained by physical system limits.

THEOREM 1 *Consider the range error dynamics* (27) *and the auxiliary nonlinear system* (28). *The follower's throttle command, $\delta_t$ given in* (33), *ensures that the trajectories of the system* (28) *remain uniformly ultimately bounded within a compact set* $\Omega_t := \left\{ (e_r, s_1^r, s_2^r, \tilde{x}_1^r, \tilde{x}_2^r) \in (\mathbb{R})^5 \,\middle|\, |e_r|^2 + |s_1^r|^2 + |s_2^r|^2 + |\tilde{x}_1^r|^2 + |\tilde{x}_2^r|^2 \leq \frac{(\epsilon_1^r)^2 + (\epsilon_2^r)^2}{w_1} \right\}$ *if the gain parameters and the filter constants satisfy* $K_0^r > \frac{(g_0^r)^2}{2} + w_1, K_1^r > \frac{(g_1^r P_r)^2}{2} + 1 + w_1, K_2^r > 1 + w_1$, *and* $k_0^r > \bar{d}_0^r, k_1^r > \bar{d}_1^r, k_2^r > \bar{d}_2^r, \frac{1}{\tau_1^r} > \frac{3}{2} + w_1, \frac{1}{\tau_2^r} > \frac{3}{2} + w_1$, *where $w_1 > 0$ is a constant.*

*Proof:*
Consider a Lyapunov function candidate, $V_r = \frac{1}{2}(e_r)^2 + \sum_{i=1}^2 \frac{1}{2}(s_i^r)^2 + \sum_{i=1}^2 \frac{1}{2}(\tilde{x}_i^r)^2$. On differentiating $V_r$ with respect to time we obtain, $\dot{V}_r = e_r \dot{e}_r + \sum_{i=1}^2 s_i^r \dot{s}_i^r + \sum_{i=1}^2 \tilde{x}_i^r \dot{\tilde{x}}_i^r$. Simplifying $\dot{V}_r$ using (28)-(32) yields

$$\begin{aligned}\dot{V}_r &= e_r\left(f_0^r + g_0^r x_{1d}^{r*} + g_0^r(s_1^r + \tilde{x}_1^r) + d_0^r\right) e_r \\ &+ s_1^r\left(f_1^r + g_1^r x_{2d}^{r*} + g_1^r P_r(s_2^r + \tilde{x}_2^r) + d_1^r\right) \\ &+ s_2^r\left(f_2^r + g_2^r u + d_2 - \dot{x}_{2c}^r\right) + \sum_{i=1}^2 \tilde{x}_i^r\left(\dot{x}_{ic}^r - \dot{x}_{id}^r\right).\end{aligned} \quad (35)$$

Letting $x_{1d}^r, x_{2d}^{r*}, u, \dot{x}_{1c}^r,$ and $\dot{x}_{2c}^r$ as proposed in (33), we obtain,

$$\begin{aligned}\dot{V}_r &= e_r\left(-K_0^r e_r - k_0^r \text{sign}(e_r) + g_0^r(s_1^r + \tilde{x}_1^r) + d_0^r\right) \\ &+ s_1^r\left(-K_1^r s_1^r - k_1^r \text{sign}(s_1^r) + g_1^r P_r(s_2^r + \tilde{x}_2^r) + d_1^r\right) \\ &+ s_2^r\left(-K_2^r s_2^r - k_2^r \text{sign}(s_2^r) + d_2\right) + \tilde{x}_1^r\left(-\frac{\tilde{x}_1^r}{\tau_1^r} - \dot{x}_{1d}^r\right) \\ &+ \tilde{x}_2^r\left(-\frac{\tilde{x}_2^r}{\tau_2^r} - \dot{x}_{2d}^r\right).\end{aligned} \quad (36)$$

Further simplification of the above expression yields,

$$\begin{aligned}\dot{V}_r &= -K_0^r e_r^2 - \left(k_0^r - |d_0^r|\right)|e_r| + e_r g_0^r(s_1^r + \tilde{x}_1^r) - \tilde{x}_1^r \frac{\tilde{x}_1^r}{\tau_1^r} \\ &- \tilde{x}_1^r \dot{x}_{1d}^r - K_1^r s_1^{r2} - \left(k_1^r - |d_1^r|\right)|s_1^r| + x_1^r g_1^r P_r(s_2^r + \tilde{x}_2^r) \\ &- K_2^r s_2^{r2} - \left(k_2^r - |d_2^r|\right)|s_2^r| - \tilde{x}_2^r \frac{\tilde{x}_2^r}{\tau_2^r} - \tilde{x}_2^r \dot{x}_{2d}^r.\end{aligned} \quad (37)$$

To bound the cross terms and disturbance-dependent components, we apply Young's inequality to obtain,

$$\dot{V}_r \leq -\left(K_0^r - \frac{(g_0^r)^2}{2}\right)e_r^2 - (k_0^r - |d_0^r|)|e_r|$$
$$- \left(K_1^r - \frac{(g_1^r P_r)^2}{2} - 1\right)s_1^{r2} - (k_1^r - |d_1^r|)|s_1^r|$$
$$- (K_2^r - 1)s_2^{r2} - (k_2^r - |d_2^r|)|s_2^r| - \tilde{x}_1^r\left(\frac{1}{\tau} - \frac{3}{2}\right)\tilde{x}_1^r$$
$$+ \frac{|\dot{x}_{1d}^r|^2}{2} - \tilde{x}_2^r\left(\frac{1}{\tau} - \frac{3}{2}\right)\tilde{x}_2^r + \frac{|\dot{x}_{2d}^r|^2}{2}, \quad (38)$$

which presents the sufficient conditions on controller parameters as in Theorem 1 to ensure all terms except the last two terms in the above expression of $\dot{V}_r$ are always negative. To show that quadratic terms dominate the positive terms in (38), we can rewrite $\dot{V}_r$ using the obtained sufficient condition as

$$\dot{V}_r \leq -\frac{w_1}{2}|e_r|^2 - \frac{w_1}{2}|s_1^r|^2 - \frac{w_1}{2}|s_2^r|^2 - \frac{w_1}{2}|\tilde{x}_1^r|^2$$
$$- \frac{w_1}{2}|\tilde{x}_2^r|^2 + \frac{|\dot{x}_{1d}^r|^2}{2} + \frac{|\dot{x}_{2d}^r|^2}{2}. \quad (39)$$

Under Assumption 1, it readily follows from the above expression that the decrement of $V_r$ is guaranteed outside a compact set $\Omega_t$. This implies that the trajectories of (28) asymptotically converge to the compact set $\Omega_t$ within which ultimate performance bounds on the errors are given as, $|e_r|, |s_1^r|, |s_2^r|, |\tilde{x}_1^r|, |\tilde{x}_2^r| \leq \sqrt{\frac{(\epsilon_1^r)^2 + (\epsilon_2^r)^2}{w_1}}$. This concludes the proof. ∎

*Remark 7* The bounds on the error variables as indicated in Theorem 1 depend on the smoothness of the follower's desired speed $x_{1d}^r$, propeller speed command $x_{2d}^r$, and the parameter $w_1$. Particularly, smoother values of $x_{1d}^r$, $x_{2d}^r$ (small $\epsilon_1^r$, $\epsilon_2^r$) and larger $w_1$ will result in a lower ultimate bound. However, selecting a very large value of $w_1$ may result in high controller gains, causing an increase in control demand that might not be practically available, and amplifying sensitivity to disturbances.

From Theorem 1, at steady-state $e_r \approx 0$ which leads to $\dot{e}_r = V_l \cos\sigma_l - V_f \cos\sigma_f = 0$ using (27), which gives us the relationship $V_f = V_l \frac{\cos\sigma_l}{\cos\sigma_f}$. If $\sigma_f \in [0, \pi/2)$, then $0 \leq \cos\sigma_f < 1$, and if $\sigma_f \in [\frac{\pi}{2}, \pi]$, then $-1 \leq \cos\sigma_f \leq 0$. This implies that the signs of $\sigma_f, \sigma_l$ are always the same since the speeds are always positive in practice. Therefore, as per Definition 3, to ensure that the follower settles behind the leader, one has to ensure that $\sigma_f \in (-\pi/2, \pi/2)$ via appropriate maneuvers.

### B. Design of the Follower's Bearing angle Controller

Let us define the followers' bearing angle errors as

$$e_\gamma = \gamma_f - \gamma_L - \sigma_{fd}^\gamma, \quad e_\chi = \chi_f - \chi_L - \sigma_{fd}^\chi, \quad (40)$$

The goal here is to nullify (40) while ensuring that the follower stays behind the leader at all times (see Definition 3). To provide the follower with sufficient maneuvering flexibility without compromising this geometric requirement, we impose explicit bounds on the bearing errors such that $|e_\gamma| < \overline{e}_\gamma$ and $|e_\chi| < \overline{e}_\chi$. These bounds guarantee that the follower's velocity orientation remains confined within a feasible sector of the LOS (see Fig. 3), thereby preserving the formation geometry and preventing loss of controllability.

Differentiating $e_\gamma, e_\chi$ in (40) with respect to time and using (3), (4), (8), (9), we obtain the dynamics of bearing angle errors as

$$\dot{e}_\gamma = -\frac{g c_{\gamma_f}}{V_f} + \frac{T_f(c_{\mu_f} s_{\alpha_f} + c_{\alpha_f} s_{\beta_f} s_{\mu_f})}{m_g V_f} - \frac{F_Y c_{\beta_f} s_{\mu_f}}{m_g V_f}$$
$$+ \frac{F_L c_{\mu_f}}{m_g V_f} - \frac{F_D s_{\beta_f} s_{\mu_f}}{m_g V_f} - \frac{V_l}{r}\left(c_{\gamma_L} s_{\gamma_l} - c_{\gamma_l} s_{\gamma_L} c_{(\chi_L - \chi_l)}\right)$$
$$+ \frac{V_f}{r}\left(c_{\gamma_L} s_{\gamma_f} - c_{\gamma_f} s_{\gamma_L} c_{(\chi_L - \chi_f)}\right), \quad (41)$$

$$\dot{e}_\chi = \frac{T_f(s_{\alpha_f} s_{\mu_f} - c_{\alpha_f} c_{\mu_f} s_{\beta_f})}{m_f V_f c_{\gamma_f}} + \frac{F_Y c_{\beta_f} c_{\mu_f}}{m_g V_f c_{\gamma_f}} + \frac{F_L s_{\mu_f}}{m_g V_f c_{\gamma_f}}$$
$$+ \frac{F_D c_{\mu_f} s_{\beta_f}}{m_f V_f c_{\gamma_f}} + \frac{V_l c_{\gamma_l} s_{(\chi_L - \chi_l)} - V_f c_{\gamma_f} s_{(\chi_L - \chi_f)}}{r c_{\gamma_L}}. \quad (42)$$

Selecting state and control vectors as $\mathbf{e}_\sigma = [e_\gamma, e_\chi]^\top$, $\mathbf{x}_1^{\sigma*} = [\alpha_f \cos\mu_f, \alpha_f \cos\mu_f]^\top$, $\mathbf{x}_1^\sigma = [\alpha_f, \beta_f, \mu_f]^\top$, $\mathbf{x}_2^\sigma = [p, q, r]^\top$, $\mathbf{u}_\sigma = [\delta_a, \delta_e, \delta_r]^\top$ and using eqs. (10) to (13), (20) to (22), (41) and (42), we can express the bearing angle errors to the control surface deflections relationship in a strict feedback form

$$\dot{\mathbf{e}}_\sigma = \mathbf{f}_0^\sigma + \mathbf{G}_0^\sigma \mathbf{x}_1^{\sigma*} + \mathbf{d}_0^\sigma, \quad (43a)$$
$$\dot{\mathbf{x}}_1^\sigma = \mathbf{f}_1^\sigma + \mathbf{G}_1^\sigma \mathbf{x}_2^\sigma + \mathbf{d}_1^\sigma, \quad (43b)$$
$$\dot{\mathbf{x}}_2^\sigma = \mathbf{f}_2^\sigma + \mathbf{G}_2^\sigma \mathbf{u}_\sigma + \mathbf{d}_2^\sigma, \quad (43c)$$

where,

$$\mathbf{f}_0^\sigma = \begin{bmatrix} V_l(\sin\gamma_L \sin\gamma_l + \cos\gamma_L \cos\gamma_l \cos(\chi_L - \chi_l)) \\ -\frac{g\cos\gamma_i}{V_i} - \frac{V_l(\cos\gamma_L \sin\gamma_l - \cos\gamma_l \sin\gamma_L \cos(\chi_L - \chi_l))}{r} \\ \frac{V_l \cos\gamma_l \sin(\chi_L - \chi_l)}{r \cos\gamma_L}, \end{bmatrix}, \quad (44)$$

$$\mathbf{G}_0^\sigma = \begin{bmatrix} s_{\gamma_L} s_{\gamma_f} + c_{\gamma_L} c_{\gamma_f} c_{\Delta\chi} & 0 & 0 \\ \frac{c_{\gamma_L} s_{\gamma_f} - c_{\gamma_f} s_{\gamma_L} c_{\Delta\chi}}{c_{\gamma_f} s_{\Delta\chi}} & \frac{\rho V_f S C_{L\alpha}}{2m} & -\frac{\rho V_f S C_{D\alpha} s_\beta}{2m} \\ \frac{c_{\gamma_f} s_{\Delta\chi}}{r c_{\gamma_L}} & \frac{\rho V_f S C_{D\alpha} s_\beta}{2m \cos\gamma} & \frac{\rho V_f S C_{L\alpha}}{2m \cos\gamma} \end{bmatrix}, \quad (45)$$

$$\mathbf{f}_1^\sigma = \begin{bmatrix} -\frac{F_L + T s_{\alpha_f} - m g c_{\gamma_f} c_{\mu_f}}{m V_f c_{\beta_f}} \\ \frac{F_y c_{\beta_f} + F_D s_{\beta_f} - T c_{\alpha_f} s_{\beta_f} + m g c_{\gamma_f} s_{\mu_f}}{m V_f} \\ -\frac{g c_{\mu_f} c_{\gamma_f} \tan\beta_f}{V_f} + \frac{F_{bx} C_1 + F_{bz} C_2 + F_{ay} C_3}{m V_t} \end{bmatrix} \quad (46)$$

$$\mathbf{G}_1^\sigma = \begin{bmatrix} -\cos\alpha_f \tan\beta_f & 1 & -\sin\alpha_f \tan\beta_f \\ \sin\alpha_f & 0 & -\cos\alpha_f \\ \frac{\cos\alpha}{\cos\beta} & 0 & \frac{\sin\alpha}{\cos\beta} \end{bmatrix} \quad (47)$$

$$\mathbf{f}_2^\sigma = \begin{bmatrix} \Gamma_1 pq - \Gamma_2 qr + \Gamma_3\left(l - C_4\right) + \Gamma_4\left(n - C_5\right) \\ \Gamma_5 pr - \Gamma_6\left(p^2 - r^2\right) + \frac{1}{J_y} m \\ \Gamma_7 pq - \Gamma_1 qr + (l - C_4) + \Gamma_8\left(n - C_5\right) \end{bmatrix} \quad (48)$$

$$\mathbf{G}_2^\sigma = bQ \begin{bmatrix} \Gamma_3 C_{l\delta_a} + \Gamma_4 C_{n\delta_a} & 0 & \Gamma_3 C_{l\delta_r} + \Gamma_4 C_{n\delta_r} \\ 0 & \frac{cC_{m\delta_e}}{bJ_y} & 0 \\ \Gamma_4 C_{l\delta_a} + \Gamma_8 C_{n\delta_a} & 0 & \Gamma_4 C_{l\delta_r} + \Gamma_8 C_{n\delta_r} \end{bmatrix} \quad (49)$$

with $\Delta\chi = \chi_f - \chi_L$, $F_{bx} = -F_{Df}\cos\alpha_f + F_{Lf}\sin\alpha_f$, $F_{bz} = F_{Lf}\cos\alpha_f + F_{Df}\sin\alpha_f$, $C_4 = C_{l\delta_a}\delta_a + C_{l\delta_r}\delta_r$, $C_5 = C_{n\delta_a}\delta_a - C_{n\delta_r}\delta_r$, and $\mathbf{d}_i^\sigma$ are the residual terms. We assume that these terms are unknown but bounded, that is, $\|\mathbf{d}_i^\sigma\| < \overline{d}_i^\sigma$, for some $\overline{d}_i^\sigma > 0 \ \forall i \in \{1, 2\}$.

To facilitate dynamic surface-based control, we define errors between the desired and filtered values of virtual controls for the system in (43) as,

$$\tilde{\mathbf{x}}_1^\sigma = \mathbf{x}_{1c}^\sigma - \mathbf{x}_{1d}^\sigma, \quad \tilde{\mathbf{x}}_2^\sigma = \mathbf{x}_{2c}^\sigma - \mathbf{x}_{2d}^\sigma, \quad (50)$$

where $\mathbf{x}_{1d}^\sigma = [\alpha_{fd}, \beta_{fd}, \mu_{fd}]^\top$ denotes the desired attitude angle, $\mathbf{x}_{1c}^\sigma = [\alpha_{fc}, \beta_{fc}, \mu_{fc}]^\top$ denotes the filtered desired attitude angles, $\mathbf{x}_{2d}^\sigma = [p_d, q_d, r_d]^\top$ represents desired body rates and $\mathbf{x}_{2c}^\sigma = [p_c, q_c, r_c]^\top$ is the filtered values of the desired body rates. Additionally, we define the error between the actual and filtered desired values as,

$$\mathbf{s}_1^\sigma = \mathbf{x}_1^\sigma - \mathbf{x}_{1c}^\sigma, \quad \mathbf{s}_2^\sigma = \mathbf{x}_2^\sigma - \mathbf{x}_{2c}^\sigma. \quad (51)$$

Note that $\mathbf{x}_{1d}^*$ is a nonlinear function of $x_{1d}$ and $x_{1c}$. Therefore, following Remark 5, we obtain a linear approximation of the deviation of the desired values from the filtered ones,

$$\mathbf{x}_1^* - \mathbf{x}_{1d}^* = \underbrace{\begin{bmatrix} 1 & 0 & 0 \\ 0 & \cos(\eta_\mu) & -\eta_\alpha \sin(\eta_\mu) \\ 0 & \sin(\eta_\mu) & \eta_\alpha \cos(\eta_\mu) \end{bmatrix}}_{\mathbf{P}} \begin{bmatrix} V_f - V_{fd} \\ \alpha_f - \alpha_{fd} \\ \mu_f - \mu_{fd} \end{bmatrix}$$
$$(52)$$

where $\eta_\mu = \mu_{fd} + \eta_1\left(\mu_f - \mu_{fd}\right)$ and $\eta_\alpha = \alpha_{fd} + \eta_2\left(\alpha_f - \alpha_{fd}\right)$, such that $0 < \eta_1 < 1$. Using eqs. (50) to (52) in (43), we have the dynamics of outer-loop error variables as

$$\dot{\mathbf{e}}_\sigma = \mathbf{f}_0^\sigma + \mathbf{G}_0^\sigma\left(\mathbf{x}_{1d}^{\sigma\,*} + \mathbf{P}\left(\mathbf{s}_1^\sigma + \tilde{\mathbf{x}}_1^\sigma\right)\right) + \mathbf{d}_0^\sigma, \quad (53)$$
$$\dot{\mathbf{x}}_1^\sigma = \mathbf{f}_1^\sigma + \mathbf{G}_1^\sigma\left(\mathbf{x}_{2d}^\sigma + \mathbf{s}_2^\sigma + \tilde{\mathbf{x}}_2^\sigma\right) + \mathbf{d}_1^\sigma. \quad (54)$$

Therefore, we now propose the follower's bearing angle control law as

$$\mathbf{u}_\sigma = \left(\mathbf{G}_2^\sigma\right)^{-1}\left(-\mathbf{f}_2^\sigma - \mathbf{K}_2^\sigma \mathbf{s}_2^\sigma - k_2^\sigma \frac{\mathbf{s}_2^\sigma}{\|\mathbf{s}_2^\sigma\|} + \dot{\mathbf{x}}_{2c}\right), \quad (55)$$

where one has the terms

$$\mathbf{x}_{1d}^{\sigma\,*} = \left(\mathbf{G}_0^\sigma\right)^{-1}\left(-\mathbf{f}_0^\sigma - k_0 \mathbf{B}_L \frac{\mathbf{e}_\sigma}{\|\mathbf{e}_\sigma\|} - \mathbf{K}_0 \mathbf{B}_L \mathbf{e}_\sigma\right) \quad (56)$$

$$\mathbf{x}_{1d}^\sigma = \begin{bmatrix} \text{sign}\left(\mathbf{x}_{1d}^{\sigma\,*}(1)\right)\sqrt{\mathbf{x}_{1d}^{\sigma\,*}(2)^2 + \mathbf{x}_{1d}^{\sigma\,*}(2)^2} \\ \arctan\left(\frac{\mathbf{x}_{1d}^{\sigma\,*}(2)}{\mathbf{x}_{1d}^{\sigma\,*}(1)}\right) \end{bmatrix}, \quad (57)$$

$$\tau_1^\sigma \dot{\mathbf{x}}_{1c}^\sigma = \mathbf{x}_{1d}^\sigma - \mathbf{x}_{1c}^\sigma, \quad \mathbf{x}_{1c}^\sigma(0) = \mathbf{x}_{1d}^\sigma(0), \quad (58)$$

$$\mathbf{x}_{2d}^\sigma = \left(\mathbf{G}_1^\sigma\right)^{-1}\left(-\mathbf{f}_1^\sigma - \mathbf{K}_1^\sigma \mathbf{s}_1^\sigma - k_1^\sigma \frac{\mathbf{s}_1^\sigma}{\|\mathbf{s}_1^\sigma\|} + \dot{\mathbf{x}}_{1c}^\sigma\right) \quad (59)$$

$$\tau_2^\sigma \dot{\mathbf{x}}_{2c}^\sigma = \mathbf{x}_{2d}^\sigma - \mathbf{x}_{2d}^\sigma, \quad \mathbf{x}_{2d}^\sigma(0) = \mathbf{x}_{2d}^\sigma(0) \quad (60)$$

with $\mathbf{K}_0^\sigma, \mathbf{K}_1^\sigma, \mathbf{K}_1^\sigma \in \mathbb{R}^{3\times3}$ are diagonal gain matrices, $k_0^\sigma$, $k_1^\sigma, k_2^\sigma \in \mathbb{R}$ are scalar gain parameters, $\tau_1^\sigma, \tau_2^\sigma \in \mathbb{R}^{3\times3}$ are the diagonal matrices denoting filter constants, and $\mathbf{B}_L = \text{diag}\left(\frac{1}{\overline{e}_\gamma^2 - e_\gamma^2}, \frac{1}{\overline{e}_\chi^2 - e_\chi^2}\right)$ is a weighting matrix.

*Assumption* 2 The time derivative of the desired virtual commands in (56) and (59) are bounded, that is, $\|\dot{\mathbf{x}}_{id}\| < \epsilon_i^\sigma$, for some $\epsilon_i^\sigma > 0, \forall i \in \{1, 2\}$.

THEOREM 2 *Consider the elevation and azimuth bearing angle error dynamics* (41)-(42) *and the auxiliary nonlinear system* (43). *The follower's surface deflection commands* (55) *will ensure that the trajectories of the system* (43) *remain uniformly ultimately bounded within a compact set* $\Omega_\sigma := \Big\{(\mathbf{e}_\sigma, \mathbf{s}_1^\sigma, \mathbf{s}_2^\sigma, \tilde{\mathbf{x}}_1^\sigma, \tilde{\mathbf{x}}_2^\sigma) \in \mathcal{D}_\sigma \Big| \|\mathbf{e}_\sigma\|^2 + \|\mathbf{s}_1^\sigma\|^2 + \|\mathbf{s}_2^\sigma\|^2 + \|\tilde{\mathbf{x}}_1^\sigma\|^2 + \|\tilde{\mathbf{x}}_2^\sigma\|^2 \leq \frac{(\epsilon_1^\sigma)^2 + (\epsilon_2^\sigma)^2}{w_2}\Big\}$, where $\mathcal{D} := (-\overline{e}_\gamma, \overline{e}_\gamma) \times (-\overline{e}_\chi, \overline{e}_\chi) \times (\mathbb{R}^3)^4$, *if the gain parameters and the filter constant satisfy* $\mathbf{K}_0^\sigma - \frac{\mathbf{G}_0^\sigma \mathbf{P} \mathbf{P}^\top \mathbf{G}_0^{\sigma\top}}{2} \geq \frac{w_2}{2} \mathbf{I}_3, \mathbf{K}_1^\sigma > \frac{\mathbf{G}_1^{\sigma\top} \mathbf{G}_1^\sigma}{2} + \left(1 + \frac{w_2}{2}\right) \mathbf{I}_3$, $\mathbf{K}_2^\sigma > \left(1 + \frac{w_2}{2}\right) \mathbf{I}_3, k_0^\sigma > \frac{\|\mathbf{d}_0\| \cdot \|\mathbf{e}_\sigma\|}{\|\mathbf{B}_L \mathbf{e}_\sigma\|}, k_1^\sigma > \|\mathbf{s}_1^\sigma\| \cdot \|\mathbf{d}_1^\sigma\|$, $k_2^\sigma > \|\mathbf{s}_2^\sigma\| \cdot \|\mathbf{d}_2^\sigma\|, (\tau_1^\sigma)^{-1}, (\tau_2^\sigma)^{-1} > \left(\frac{3}{2} + w_2\right) \mathbf{I}_3$ *with* $w_2 \in \mathbb{R}_{>0}$ *being a constant.*

*Proof:*
Consider a Lyapunov function candidate,

$$V_\sigma = \frac{1}{2}\log\left(\frac{\overline{e}_\gamma^2}{\overline{e}_\gamma^2 - e_\gamma^2}\right) + \frac{1}{2}\log\left(\frac{\overline{e}_\chi^2}{\overline{e}_\chi^2 - e_\chi^2}\right)$$
$$+ \sum_{i=1}^{2}\left((\tilde{\mathbf{x}}_i^\sigma)^\top \tilde{\mathbf{x}}_i^\sigma + \frac{1}{2}(\mathbf{s}_i^\sigma)^\top \mathbf{s}_i^\sigma\right), \quad (61)$$

which is radially unbounded within the constraint set defined by $|e_\gamma| < \overline{e}_\gamma$ and $|e_\chi| < \overline{e}_\chi$. Differentiating $V_\sigma$ with respect to time, we obtain

$$\dot{V}_\sigma = \frac{e_\chi \dot{e}_\chi}{\overline{e}_\chi^2 - e_\chi^2} + \frac{e_\chi \dot{e}_\chi}{\overline{e}_\chi^2 - e_\chi^2} + \sum_{i=1}^{2}\left((\tilde{\mathbf{x}}_\mathbf{i}^\sigma)^\top \tilde{\mathbf{x}}_\mathbf{i}^\sigma + \frac{1}{2}(\mathbf{s}_\mathbf{i}^\sigma)^\top \mathbf{s}_\mathbf{i}^\sigma\right)$$
$$= \mathbf{B}_L \mathbf{e}_\sigma^\top \dot{\mathbf{e}}_\sigma + \sum_{i=1}^{2}\left((\tilde{x}_i^\sigma)^\top \tilde{\mathbf{x}}_\mathbf{i}^\sigma + \frac{1}{2}(\mathbf{s}_i^\sigma)^\top \dot{\mathbf{s}}_\mathbf{i}^\sigma\right) \quad (62)$$

Substituting (53), (54) and (43c) in the above equation, we obtain

$$\dot{V}_\sigma = \mathbf{e}_\sigma^\top \mathbf{B}_L\left(\mathbf{f}_0^\sigma + \mathbf{G}_0^\sigma\left(\mathbf{x}_1^{\sigma\,*} + \mathbf{P}\left(\mathbf{s}_1^\sigma + \tilde{\mathbf{x}}_1^\sigma\right)\right) + \mathbf{d}_0\right) \quad (63)$$
$$+ (\mathbf{s}_1^\sigma)^\top\left(\mathbf{f}_1^\sigma + \mathbf{G}_1^\sigma\left(\mathbf{x}_{2d}^\sigma + \mathbf{s}_2^\sigma + \tilde{\mathbf{x}}_2^\sigma\right) + \mathbf{d}_1^\sigma\right)$$
$$+ (\mathbf{s}_2^\sigma)^\top\left(\mathbf{f}_2^\sigma + \mathbf{G}_2^\sigma \mathbf{u}_\sigma + \mathbf{d}_2^\sigma - \dot{\mathbf{x}}_{2c}^\sigma\right)$$
$$+ (\tilde{\mathbf{x}}_1^\sigma)^\top\left(\dot{\mathbf{x}}_{1c}^\sigma - \dot{\mathbf{x}}_{1d}^\sigma\right) + (\tilde{\mathbf{x}}_2^\sigma)^\top\left(\dot{\mathbf{x}}_{2c}^\sigma - \dot{\mathbf{x}}_{2d}^\sigma\right). \quad (64)$$

Now using the proposed control law (55), the above expression reduces to

$$\dot{V}_\sigma = \mathbf{B}_L \mathbf{e}_\sigma^\top \left[ \frac{-k_0 \mathbf{B}_L \mathbf{e}_\sigma}{\|\mathbf{e}_\sigma\|} - \mathbf{K}_0^\sigma \mathbf{B}_L \mathbf{e}_\sigma + \mathbf{G}_0 \mathbf{P}(\mathbf{s}_1^\sigma + \tilde{\mathbf{x}}_1^\sigma) \right.$$
$$\left. + \mathbf{d}_0^\sigma \right] + (\tilde{\mathbf{x}}_1^\sigma)^\top \left( -(\tau_1^\sigma)^{-1} \tilde{\mathbf{x}}_1^\sigma - \dot{\mathbf{x}}_{1d}^\sigma \right)$$
$$+ (\mathbf{s}_1^\sigma)^\top \left( -\mathbf{K}_1^\sigma \mathbf{s}_1^\sigma - k_1^\sigma \frac{\mathbf{s}_1^\sigma}{\|\mathbf{s}_1^\sigma\|} + \mathbf{d}_1^\sigma + \mathbf{G}_1^\sigma \left( \mathbf{s}_2^\sigma + \tilde{\mathbf{x}}_2^\sigma \right) \right)$$
$$+ (\mathbf{s}_2^\sigma)^\top \left( -\mathbf{K}_2^\sigma \mathbf{s}_2^\sigma - k_2^\sigma \frac{\mathbf{s}_2^\sigma}{\|\mathbf{s}_2^\sigma\|} + \mathbf{d}_2^\sigma \right)$$
$$+ (\tilde{\mathbf{x}}_2^\sigma)^\top \left( -(\tau_2^\sigma)^{-1} \tilde{\mathbf{x}}_2^\sigma - \dot{\mathbf{x}}_{2d}^\sigma \right),$$

which, on further simplification, leads to

$$\dot{V}_\sigma \leq -\mathbf{e}_\sigma^\top \mathbf{B}_L^\top \left( \mathbf{K}_0 - \frac{\mathbf{G}_0^\sigma \mathbf{P}\mathbf{P}^\top \mathbf{G}_0^{\sigma\top}}{2} \right) \mathbf{B}_L \mathbf{e}_\sigma$$
$$- \|\mathbf{B}_L \mathbf{e}_\sigma\| \left( \frac{k_0}{\|\mathbf{e}_\sigma\|} \|\mathbf{B}_L \mathbf{e}_\sigma\| - \|\mathbf{d}_0^\sigma\| \right)$$
$$- (\mathbf{s}_1^\sigma)^\top \left( \mathbf{K}_1^\sigma - \frac{\mathbf{G}_1^{\sigma\top} \mathbf{G}_1^\sigma}{2} - \mathbf{I}_3 \right) \mathbf{s}_1^\sigma$$
$$- \left( \frac{k_1^\sigma}{\|\mathbf{s}_1^\sigma\|} - \|\mathbf{d}_1^\sigma\| \right) \|\mathbf{s}_1^\sigma\| - (\mathbf{s}_2^\sigma)^\top \left( \mathbf{K}_2^\sigma - \mathbf{I}_3 \right) \mathbf{s}_2^\sigma$$
$$- \left( \frac{k_2^\sigma}{\|\mathbf{s}_2^\sigma\|} - \|\mathbf{d}_2^\sigma\| \right) \|\mathbf{s}_2^\sigma\| - (\tilde{\mathbf{x}}_1^\sigma)^\top \left( (\tau_1^\sigma)^{-1} - \frac{3}{2}\mathbf{I}_3 \right) \tilde{\mathbf{x}}_1^\sigma$$
$$- (\tilde{\mathbf{x}}_2^\sigma)^\top \left( (\tau_2^\sigma)^{-1} - \frac{3}{2}\mathbf{I}_3 \right) \tilde{\mathbf{x}}_2^\sigma + \frac{\|\dot{\mathbf{x}}_{1d}^\sigma\|^2}{2} + \frac{\|\dot{\mathbf{x}}_{2d}^\sigma\|^2}{2},$$
(65)

yielding the sufficient conditions presented in Theorem 2 on the controller gains and filter constants to ensure that the quadratic and disturbance-related terms are negative definite in (65).

To demonstrate that the negative terms in (65) dominate the positive term, we can further simplify $\dot{V}_\sigma$ using the sufficient condition to obtain

$$\dot{V}_\sigma \leq -\frac{w_2}{2} \|\mathbf{B}_L e_\sigma\|^2 - \frac{w_2}{2} \sum_{i=1}^{2} \left( \|\mathbf{s}_i^\sigma\|^2 + \|\tilde{\mathbf{x}}_i^\sigma\|^2 \right)$$
$$+ \frac{\|\dot{\mathbf{x}}_{1d}^\sigma\|^2}{2} + \frac{\|\dot{\mathbf{x}}_{2d}^\sigma\|^2}{2}. \quad (66)$$

Under Assumption 2, it follows from the above equation that the decrement of $V_\sigma$ is guaranteed outside the compact set $\Omega_\sigma$. This implies that closed-loop trajectories of the system (43) are uniformly ultimately bounded with ultimate performance bounds

$$\sqrt{\lambda_b} \|\mathbf{e}_\sigma\|, \|\mathbf{s}_1^\sigma\|, \|\mathbf{s}_2^\sigma\|, \|\tilde{\mathbf{x}}_1^\sigma\|, \|\tilde{\mathbf{x}}_2^\sigma\| \leq \sqrt{\frac{(\epsilon_1^\sigma)^2 + (\epsilon_2^\sigma)^2}{w_2}} \quad (67)$$

where $\lambda_b = \lambda_{\min}\left(\mathbf{B}_L^\top \mathbf{B}_L\right)$ denotes the smallest eigenvalue of the matrix $\mathbf{B}_L^\top \mathbf{B}_L$. This concludes the proof. ■

*Remark* 8 Since $V_\sigma$ is positive definite and radially unbounded within $\mathcal{D}$, and its derivative is always negative definite except at the origin, it follows from Lemma 1 that all system trajectories starting in $\mathcal{D}$ will remain in $\mathcal{D}$ for all future times. This implies that elevation and azimuth bearing angle errors will remain bounded as $e_\gamma(t) \in (-\bar{e}_\gamma, \bar{e}_\gamma)$ and $e_\chi(t) \in (-\bar{e}_\chi, \bar{e}_\chi)$, respectively.

*Remark* 9 From (67), we can infer that a smaller variation in $\mathbf{x}_{1d}^\sigma, \mathbf{x}_{2d}^\sigma$ leads to tighter bounds on the error variables, since $\epsilon_1^\sigma, \epsilon_2^\sigma$ depend on the smoothness of the desired body rates. Further, the parameter $w_2$ directly determines the size of the ultimate bound in $\Omega_\sigma$. A larger value of $w_2$ will yield smaller bounds on the error variables, improving the steady state tracking accuracy, but an excessively large $w_3$ may lead to aggressive control action. Therefore, $w_3$ should be selected to achieve an appropriate balance between the tracking performance and control effort.

*Remark* 10 On analyzing the term $\cos \sigma_f$ by setting $\gamma_f = \gamma_L + \sigma_{fd}^\gamma + e_\gamma$, $\chi_f - \chi_L = \sigma_{fd}^\chi + e_\chi$ and using trigonometric identities, we have, $\cos \sigma_f = \frac{1}{2}\Big( \left(1 + \cos(\Delta_\chi)\right) \cos \sigma_f^\gamma + \left(\cos(\Delta_\chi) - 1\right) \cos(2\gamma_L + \Delta_\gamma) \Big)$, where $\Delta_\gamma = \sigma_{fd}^\gamma + e_\gamma$ and $\Delta_\chi = \sigma_{fd}^\chi + e_\chi$. Since $\cos(2\gamma_L + \Delta_\gamma) \in [-1, 1]$, we have the lower bound on the term $\cos \sigma_f$ as $\cos \sigma_f > \frac{1}{2}\Big( \left(1 + \cos \Delta_\chi\right) \cos \Delta_\gamma + \left(\cos \Delta_\chi - 1\right) \Big)$. On imposing the condition of staying behind following Definition 3 (i.e., $\cos \sigma_f > 0$) on the lower bound of $\cos \sigma_f$, we obtain the inequality

$$\cos\left(\sigma_{fd}^\gamma + e_\gamma\right) > \tan^2\left(\frac{\sigma_{fd}^\chi + e_\chi}{2}\right) \quad (68)$$

which provides a condition on the elevation and azimuth bearing angles to ensure that the follower always stays behind the leader.

To further ensure that the above condition is feasible and using the maximum bounds on the bearing angle errors, we obtain $|\sigma_{fd}^\chi| + \bar{e}_\chi < \pi/2$ and $|\sigma_{fd}^\gamma| < \cos^{-1}\left(\tan^2\left(\frac{\sigma_{fd}^\chi + |e_\chi|}{2}\right)\right) - \bar{e}_\gamma$. Therefore, to ensure that the follower stays behind the leader, the desired values of follower's bearing angles and their bounds should satisfy, $\sigma_{fd}^\chi \in (-\pi/2 + \bar{e}_\chi, \pi/2 - \bar{e}_\chi)$, $\bar{e}_\chi \in (0, \pi/2)$, $\sigma_{fd}^\gamma \in (-\zeta^* + \bar{e}_\gamma, \zeta^* - \bar{e}_\gamma)$ and $\bar{e}_\gamma \in (0, \zeta^*)$, where $\zeta^* = \cos^{-1}\left(\tan^2\left(\frac{\sigma_d^\chi + |e_\chi|}{2}\right)\right)$.

## IV. Simulations

We now present simulation results to demonstrate the performance of the proposed IGC approach in relational maneuvering of a leader-follower multivehicle system for flexible formation under different leaders' maneuvers and followers' initial conditions. The follower's dynamics are modeled using the Aerosonde UAV, whose parameter values are provided in Table I [33]. The architecture of the proposed scheme is presented in Fig. 4. The control parameters for all the simulations are selected as follows $K_0^r = 0.2, K_1^r = 0.6, K_2^r = 1.5 \ k_0^r = 0.1, k_0^r = 0.3, k_0^r =$

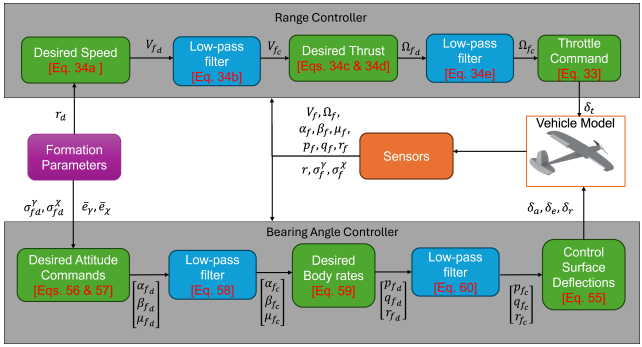

Fig. 4: IGC flexible formation control architecture.

$0.6$, $\mathbf{K}_0^\sigma = \text{diag}(0.3, 0.2)$, $\mathbf{K}_1^\sigma = \text{diag}(1.2, 1.2, 1.2)$, $\mathbf{K}_2^\sigma = \text{diag}(1.5, 1.5, 1.5)$, $k_0^\sigma = 0.3, k_1^\sigma = 5, k_2^\sigma = 2$, $\tau_1^r = \tau_2^r = 0.1$, and $\boldsymbol{\tau}_1^\sigma = \boldsymbol{\tau}_2^\sigma = \text{diag}(0.2, 0.2, 0.2)$.

TABLE I: Physical parameters and aerodynamic coefficients of the Aerosonde UAV.

| Physical Parameters | | Aerodynamic Coefficients | |
|---|---|---|---|
| Symbol | Value | Symbol | Value |
| $m_g$ | 11 kg | $C_{L0}$ | 0.28 |
| $S$ | 0.55 m$^2$ | $C_{D0}$ | 0.03 |
| $b$ | 2.89 m | $C_{L\alpha}$ | 3.45 |
| $c$ | 0.18 m | $C_{D\alpha}$ | 0.30 |
| $J_x$ | 0.8244 kg m$^2$ | $C_{l\delta_a}$ | 0.17 |
| $J_y$ | 1.135 kg m$^2$ | $C_{n\delta_a}$ | $-0.011$ |
| $J_z$ | 1.759 kg m$^2$ | $C_{l\delta_r}$ | 0.0024 |
| $J_{xz}$ | 0.1204 kg m$^2$ | $C_{n\delta_r}$ | $-0.069$ |
| $V_{\max}$ | 44.4 V | $C_{Q0}$ | 0.00523 |
| $D_P$ | 0.508 m | $C_{Q1}$ | 0.00497 |
| $K_V$ | 0.0659 V·s/rad | $C_{Q2}$ | $-0.01664$ |
| $K_Q$ | 0.0659 N·m | $C_{T0}$ | 0.09357 |
| $R$ | 0.042 Ω | $C_{T1}$ | $-0.06044$ |
| | | $C_{T2}$ | $-0.1079$ |

For the first set of results, the leader starting at $\mathbf{p}_l = [100, 100, -1000]^\top$ m executes an ascending loiter maneuver at a constant speed $V_l = 25$ m/s, with a fixed flight path angle $\gamma_l = 10°$ and fixed heading angle rate given as $\dot\chi_l = 0.1$ rad/s. The followers starts at $\mathbf{p}_f = [0, 0, -1050]^\top$ m with flight path and heading angles, $\gamma_f = 0°$ and $\chi_f = 0°$. The desired formation parameters are selected as $r_d = 50$ m, $\sigma_{fd}^\gamma = \sigma_{fd}^\chi = 0°$, $\bar{e}_\chi = 90°$ and $\bar{e}_\gamma = 80°$.

Fig. 5 depicts the leader's ascending loiter, where Fig. 5a shows the follower's trajectory converging to a small neighborhood near the desired range from the leader while remaining behind/below the latter. It is observed that the follower moves on a smaller loiter circle than the leader, demonstrating the anticipatory nature in followers' behavior by taking a shorter path rather than exactly following the trajectory traced by the leader. Fig. 5b illustrates the followers' range and bearing angle errors that converge to a small neighborhood near zero, with the bearing angle errors remaining in the predefined bounds denoted by the dotted line in the elevation and azimuth error plots. Figs. 5c and 5d presents the profiles of the internal control variables of the IGC range error and bearing angle error systems that remain smooth throughout the maneuver.

To illustrate the robustness of the IGC law, Fig. 6 illustrates the performance when the leader executes a *3D Lazy-8* maneuver at constant speed $V_l = 25$ m/s following angular speed profiles given as $\dot\gamma_l(t) = \frac{\sin(t/10)}{100}$ and $\dot\chi_l(t) = \frac{\sin(t/20)}{12 \cos\gamma_l}$. The initial conditions for the leader remains the same as in the ascending loiter case, while the follower starts at different positions (all in m) given as $F_1:[0,0,-970]^\top$, $F_2:[0,200,-970]^\top$, $F_3:[0,0,-1050]^\top$, and $F_4:[0,200,-1030]^\top$ with heading and flight path angle as $\gamma_f = 0°$ and $\chi_f = 0°$. The desired range is set as $r_d = 50$ m for all the followers, and the desired bearing angles for the followers are selected, correspondingly, as, $F_1$: $\sigma_{fd}^\gamma = \sigma_{fd}^\chi = -30°$, $F_2$: $\sigma_{fd}^\gamma = -30°, \sigma_{fd}^\chi = 30°$, $F_3$: $\sigma_{fd}^\gamma = 30°, \sigma_{fd}^\chi = -30°$, and $F_4$: $\sigma_{fd}^\gamma = \sigma_{fd}^\chi = 30°$. Furthermore, the bounds on the bearing angles are selected as $\bar{e}_\chi = 35°$ and $\bar{e}_\gamma = 30°$. In Fig. 6a, it is observed that followers starting from different initial conditions converge arbitrarily close to the desired range from the leader, with the formation parameters $\sigma_{fd}^\gamma, \sigma_{fd}^\chi$ determining the region behind the leader to which the follower converges to. The bearing angle errors remain bounded within the predefined bounds as shown in Fig. 6b, where the dotted lines represent the desired bounds. Fig. 6d compares the follower's speed, flight path angle, and heading of the follower with respect to the leader, where it is observed that when the leader's heading angle or flight path angle increases, the follower's heading and flight path angle remain less than that of the leader. However, when the leader's heading angle or flight path angle decreases, the follower's heading and flight path angle remain more than that of the leader. This behavior reflects the follower's anticipatory maneuvers, cutting inside the leader's turn or taking a wider path as needed in both azimuth and pitch planes, allowing greater flexibility in formation by enabling mismatch in leader-follower speed, flight path, and heading angles, that is, $V_f \neq V_l, \gamma_f \neq \gamma_l$ and $\chi_f \neq \chi_l$ while staying in formation.

## V. Conclusions

In this paper, we developed an integrated guidance and control (IGC) framework for a fixed-wing UAV to realize a 3D flexible leader-follower formation. In particular, we introduced the fixed elevation and azimuth bearing angle formation (FEAM) scheme that leads the follower to maintain a fixed distance and fixed bearing angles with respect to the leader while ensuring predefined constraints on the bearing angles. The proposed IGC framework integrates aerodynamic surface dynamics and nonlinear propeller-motor characteristics within the formation con-troller design, producing a physically realistic controller, while Lyapunov barrier functions ensured that safety-critical bearing constraints are respected throughout the maneuver. Our results demonstrate that the flexibility in formation broadens the feasible set of follower positions for the follower and leads to anticipatory behavior, where the follower naturally adjusts its trajectory to accommodate aggressive leader maneuvers. The IGC design achieves the formation objectives through physically realizable inputs by coupling the guidance law with the actual vehicle dynamics, yielding smoother control actions, greater robustness to dynamic variations, and more realistic flight behavior than kinematic or simplified IGC methods. Future work will focus on extending the proposed framework to multi-follower scenarios, experimental validation on hardware platforms, and incorporating environmental uncertainties such as wind and adversarial disturbances to further enhance robustness in realistic mission settings.

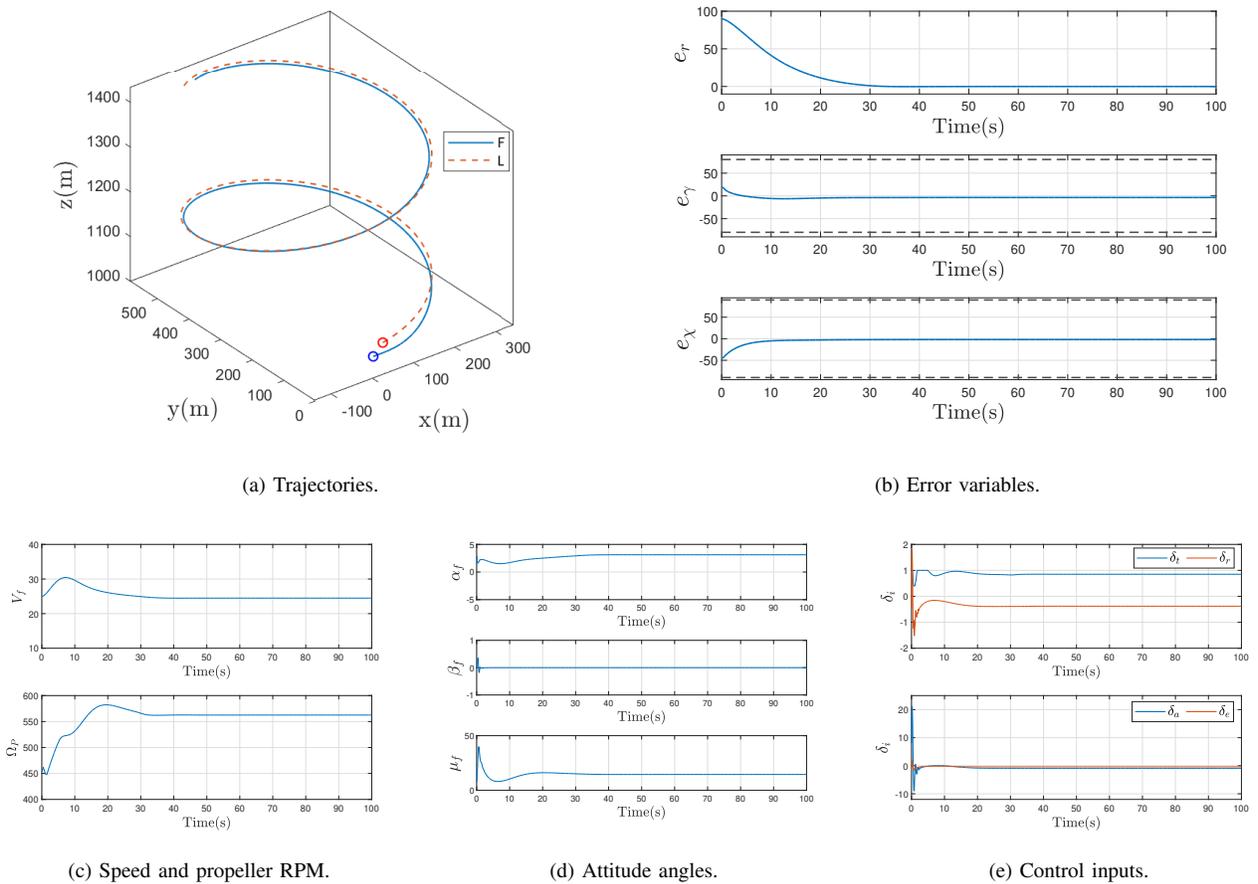

(a) Trajectories.

(b) Error variables.

(c) Speed and propeller RPM.

(d) Attitude angles.

(e) Control inputs.

Fig. 5: Performance of FEAM scheme for the leader's ascending loiter maneuver.

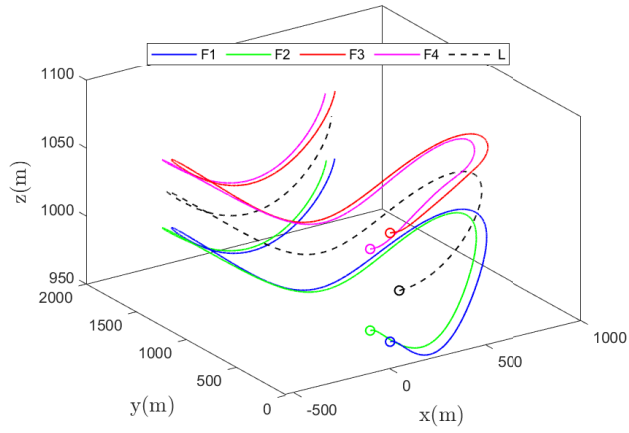

(a) Trajectories.

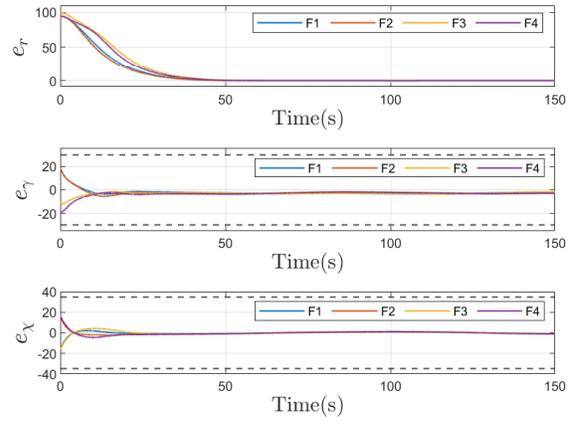

(b) Error variables.

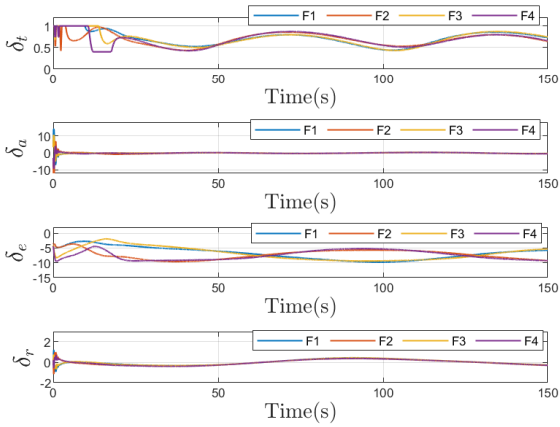

(c) Control inputs.

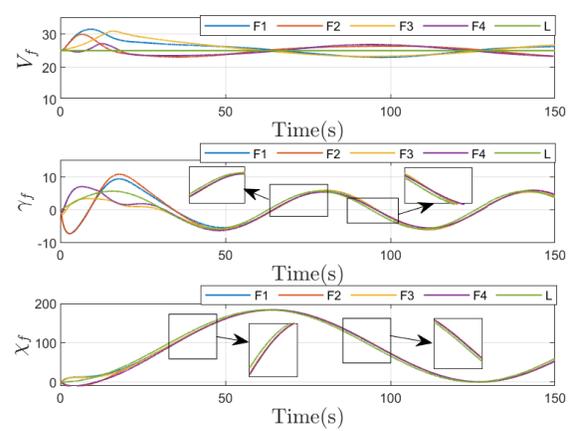

(d) Flight path and heading angle.

Fig. 6: Comparison of followers' performance for the leader's Lazy 8 maneuver.